\documentclass[referee,useAMS,usenatbib,usegraphicx]{mn2e}
\usepackage{amssymb}
\def\etal{\textit{et al.}\ }

\def\pmi{$\pm$}

\title[Optical polarimetry of NGC 654]{Optical polarimetric study of open clusters: Distribution of Interstellar matter towards NGC 654}
\author[B.J. Medhi et al.]
{Biman J. Medhi$^{1}$\thanks{E-mail: biman@aries.ernet.in},
Maheswar. G$^{1,2}$, J. C. Pandey$^{1}$, T. S. Kumar$^{1}$ and Ram Sagar$^{1}$\\
$^{1}$Aryabhatta Research Institute of Observational Sciences, Manora Peak, Nainital - 263 129, India\\
$^{2}$Korea Astronomy and Space Science Institute 61-1, Hwaam-dong, Yuseong-gu, Daejeon, Republic of Korea 305-348\\}

\begin{document}

\date{}

\pagerange{\pageref{firstpage}--\pageref{lastpage}} \pubyear{2007}

\maketitle
\label{firstpage}
\begin{abstract}
We present new $B, V$  and $R$ linear polarimetric observations for 61
stars towards  the region of the  young open cluster NGC  654. In this
study we  found evidence for  the presence of  at least two  layers of
dust along the line of sight to the cluster.  The distances to the two
dust layers are estimated to be $\sim200$ pc and $\sim1$ kpc which are
located much closer to the Sun than the cluster ($\sim2.4$ kpc).  Both
the  dust layers have  their local  magnetic field  orientation nearly
parallel to the direction of  the Galactic plane.  The foreground dust
layer  is  found to  have  a ring  morphology  with  the central  hole
coinciding with the center of  the cluster. The foreground dust grains
are  suggested  to  be   mainly  responsible  for  both  the  observed
differential reddening and the polarization towards the cluster.
\end{abstract}

\begin{keywords}
polarization- dust, extinction- open clusters and associations:individual: NGC 654
\end{keywords}

\section{Introduction}
The wavelength dependence  of interstellar extinction and polarization
provides  constraints on the  characteristics of  interstellar grains.
Interstellar polarization strongly  varies with wavelength (Serkowski,
Mathewson  \& Ford  1975;  Wilking \etal  1980).   In particular,  the
wavelength of  maximum interstellar polarization  ($\lambda_{max}$) is
thought to  be related to the  total-to-selective extinction ($R_{V}$)
as $R_{V}=(5.6\pm0.3)\lambda_{max}$ (Whittet  \& van Breda 1978).  The
polarization  of background starlight  has been  used for  nearly five
decades  to probe  the magnetic  field direction  in  the interstellar
medium (ISM).  The  observed polarization is believed to  be caused by
the dichroic  extinction of  background starlight passing  through the
concentrations of aligned and elongated  dust grains along the line of
sight. Although there is no general consensus on which is the dominant
grain  alignment mechanism (Lazarian,  Goodman \&  Myers 1997),  it is
generally believed that the elongated grains tend to become aligned to
the  local magnetic  field with  their shortest  axis parallel  to the
field.   For this  orientation,  the observed  polarization vector  is
parallel  to the plane-of-sky  projection of  a line-of-sight-averaged
magnetic field (Davis \& Greenstein 1951).

Young  open  clusters  are  very  good candidates  to  carry  out  the
polarimetric  observations  because of  available  knowledge on  their
physical parameters like  distance, membership probability $(M_p)$ and
colour  excess  $(E(B-V))$.   This  would  enable us  to  undertake  a
meaningful study of foreground interstellar dust.
  
As  a part  of an  observational programme  to carry  out polarimetric
observations  of  young  open  clusters towards  anti-galactic  center
direction  ($l  =  120^{\circ}  -  180^{\circ}$)  to  investigate  the
properties like  magnetic field orientation,  $\lambda_{max}$, maximum
polarization $(P_{max})$, etc., we  observed two young open cluster IC
1805 (Medhi  et al.  2007) and NGC  654. In this paper  we present the
results obtained  from our study on  NGC 654.  The  young open cluster
NGC    654    $(R.A.\     (J2000):01^{h}    44^{m}    00^{s},\    Dec\
(J2000):+61^{\circ}\    53^{m}\     06^{s};    \    l=129.08^{\circ},\
b=-00.36^{\circ})$ in Cassiopeia has been classified as Trumpler class
II2r by  Lyng\.{a}\ (1984).  The  post-main-sequence stars in  NGC 654
reveal  an  age  of  $\sim10-30$  Myr for  the  cluster,  whereas  the
pre-main-sequence stars  indicate an age of $\sim1-10$  Myr (Pandey et
al. 2005).   The value  of $E(B-V)$ across  the cluster varies  in the
range $\sim0.7-1.2$ mag  (Joshi \& Sagar 1983; Phelps  \& Janes 1994).
A distance modulus of  $14.7\pm0.10$ which corresponds to $2.41\pm0.11
$  kpc (for  a  normal reddening  law)  is estimated  for the  cluster
(Pandey et al. 2005).

First polarimetric measurements of the stars towards NGC 654 were made
by Samson (1976) with plates  using IIa0 emulsion (IIa0 emulsion has a
red cutoff  similar to  Johnson B band).  Samson (1976)  found certain
stars to  be very  conspicuous because they  differed in  magnitude or
direction of  polarization, or both,  from the surrounding  stars. The
star $\#37$ for e.g., was  unusual in both the magnitude and direction
of  polarization. He suggested  a dust  cloud partially  obscuring the
area to be the reason. The general direction of polarization was found
to be  parallel to the  Galactic magnetic field.  Recently,  Pandey et
al.(2005) observed  7 stars  in the  vicinity of NGC  654. Of  these 7
stars, 4  of them with higher membership  probability ($M_{P}$) showed
relatively  large polarization (3.03  to 4.47\%)  and a  mean position
angle $PA_{V}\simeq 94^{\circ}$ in V filter.  The two non-member stars
showed   relatively  small   polarization  ($\sim2\%$)   and   a  mean
$PA_{V}\simeq105^{\circ}$. The remaining  star \# 57 with $M_{P}=0.90$
(Stone 1977)  was identified as a  non-member (Sagar \&  Yu 1989) also
showed  values  (2\%  \&   $102^{\circ}$)  similar  to  those  of  two
non-member stars.   From their multi-wavelength  study, they concluded
that the  dust grains associated with  NGC 654 are  smaller than those
associated with the general interstellar medium.

In  this paper, we  present the  results of  polarimetric measurements
made for 61 stars in $B,  V$ and $R$ photometric bands towards NGC 654
(brighter than $V\simeq 17$ mag). Of the 61 stars observed, 8 of them
have high  membership probability ($M_{P}  \geq 0.7$).  The  paper is
organized  as following:  in section  2 we  present  observations; the
results and discussion are presented in  section 3 and in section 4 we
conclude with a summary.

\section{Observations}

\begin{table*}
\centering
\caption{Observed polarized and unpolarized standard stars}\label{std_obs}
\begin{tabular}{lllllll}
\hline
\multicolumn{5}{|c|}{Polarized Standard}&\multicolumn{2}{c}{Unpolarized Standard}\\
\hline
Filter&$P\pm\epsilon(\%)$ &  $\theta \pm \epsilon(^\circ)$ &  $P\pm\epsilon(\%)$ & $\theta \pm\epsilon(^\circ)$ &\ \ \ \ $q(\%)$ &\ \ \ \ \ $u(\%)$ \\
\hline
& \multicolumn{2}{c}{Schmidt \& Elston (1992)}&\multicolumn{2}{c}{This work}&\multicolumn{2}{c}{This work}\\
\hline
\multicolumn{5}{|c|}{\underline{Hiltner-960}}                            &\multicolumn{2}{|c|}{\underline{HD21447}} \\
B & $5.72\pm0.06$ & $55.06\pm 0.31$  & $5.62\pm 0.20$&$  54.65 \pm 1.04$ & \ \ \ \ \ 0.019  &\ \ \ \ \ \ 0.011  \\
V & $5.66\pm0.02$ & $54.79\pm 0.11$  & $5.70\pm 0.14$&$  53.37 \pm 0.08$ & \ \ \ \ \ 0.037  &\ \ \ \ -\ 0.031 \\
R & $5.21\pm0.03$ & $54.54\pm 0.16$  & $5.20\pm 0.06$&$  54.80 \pm 0.38$ & \ \ \ -\ 0.035  &\ \ \ \ -\ 0.039 \\
\multicolumn{5}{|c|}{\underline{HD 204827}}                              &\multicolumn{2}{|c|}{\underline{HD12021}} \\
B & $5.65\pm0.02$ & $58.20\pm 0.11$  & $5.72\pm 0.09$&$  58.60 \pm 0.49$ & \ \ \ -\ 0.108  & \ \ \ \ \ \ 0.071 \\
V & $5.32\pm0.02$ & $58.73\pm 0.08$  & $5.35\pm 0.03$&$  60.10 \pm 0.20$ & \ \ \ \ \ 0.042 & \ \ \ \ -\ 0.045 \\
R & $4.89\pm0.03$ & $59.10\pm 0.17$  & $4.91\pm 0.20$&$  58.89 \pm 1.20$ & \ \ \ \ \ 0.020 & \ \ \ \ \ \ 0.031 \\
\multicolumn{5}{|c|}{\underline{BD+64$^{\circ}$106}}                     &\multicolumn{2}{|c|}{\underline{HD14069}} \\
B & $5.51\pm0.09$ & $97.15\pm 0.47$  & $5.46\pm 0.10$&$  99.40 \pm 0.50$ &\ \ \ \ \ 0.138 &\ \ \ \ -\ 0.010 \\
V & $5.69\pm0.04$ & $96.63\pm 0.18$  & $5.48\pm 0.11$&$  97.09 \pm 0.12$ &\ \ \ \ \ 0.021 &\ \ \ \ \ \  0.018 \\
R & $5.15\pm0.10$ & $96.74\pm 0.54$  & $5.20\pm 0.02$&$  97.35 \pm 0.18$ &\ \ \ \ \ 0.010 &\ \ \ \ -\ 0.014 \\
\multicolumn{5}{|c|}{\underline{HD 19820}}                               &\multicolumn{2}{|c|}{\underline{G191B2B}} \\
B & $4.70\pm0.04$ & $115.70\pm 0.22$ & $4.81\pm 0.20$&$ 113.49 \pm 0.19$ &\ \ \ \ \ 0.072 &\ \ \ \ -\ 0.059 \\
V & $4.79\pm0.03$ & $114.93\pm 0.17$ & $4.91\pm 0.10$&$ 114.55 \pm 0.20$ &\ \ \ -\ 0.022 &\ \ \ \ -\ 0.041 \\
R & $4.53\pm0.03$ & $114.46\pm 0.17$ & $4.70\pm 0.13$&$ 113.88 \pm 0.21$ &\ \ \ -\ 0.036 &\ \ \ \ \ \ 0.027 \\
\hline
\end{tabular}
\end{table*}

The  optical  imaging  polarimetric  observations of  the  two  fields
(centered  at $R.A.:  01^{h} 43^{m}  19^{s},\ Dec:  +61^{\circ} 51^{m}
31^{s}\  and \ R.A.:  01^{h} 43^{m}  56^{s},\ Dec:  +61^{\circ} 57^{m}
48^{s}$)  in NGC  654  were  carried out  on  27$^{th}$ and  28$^{th}$
December, 2006  using ARIES  Imaging Polarimeter (AIMPOL;  Medhi \etal
2007, Rautela, Joshi  \& Pandey 2004) mounted on  the Cassegrain focus
of the  104-cm Sampurnanand telescope  of ARIES, Nainital in  $B$, $V$
and $R$  ($\lambda_{B_{eff}}$=0.440 $\mu m$, $\lambda_{V_{eff}}$=0.550
$\mu m$ and $\lambda_{R_{eff}}$=0.660 $\mu m$) photometric bands.  The
imaging was done by using a TK $1024\times 1024$ pixel$^2$ CCD camera.
Each pixel  of the CCD  corresponds to $1.7$  arcsec and the  field of
view is $\sim 8$ arcmin diameter  on the sky.  The FWHM of the stellar
image varies from $2$  to $3 \ pixel$. The read out  noise and gain of
the CCD are  7.0 $e^-$ and 11.98 $e^-$/ADU,  respectively.  The fluxes
for all of  our programme stars were extracted  by aperture photometry
after  the bias  subtraction in  the standard  manner using  IRAF. The
detail descriptions about the  AIMPOL, data reduction and calculations
of polarization, position angle are given in Medhi \etal (2007).

Standard stars  for null  polarization and for  the zero point  of the
polarization  position  angle  were   taken  from  Schmidt  \&  Elston
(1992). The observed degree of polarization ($P\%$) and position angle
($\theta$) for  the polarized  standard stars and  their corresponding
values   from  Schmidt   \&   Elston  (1992)   are   given  in   Table
\ref{std_obs}. The observed  values of $P\%$ and $\theta$  are in good
agreement  with those  given in  Schmidt \&  Elston (1992)  within the
observational errors. The  observed normalized stokes parameters $q\%$
and $u\%$ for standard unpolarized  stars (Schmidt \& Elston 1992) are
also given in Table  \ref{std_obs}.  The average value of instrumental
polarization  is  found  to   be  $\sim  0.04\%$.   

The ordinary and extraordinary images  of each source in the CCD frame
is separated by $27~pixels$ along the north-south direction on the sky
plane. Due to  the lack of a grid, placed to  avoid the overlapping of
ordinary image of one source with the extraordinary of an adjacent one
located $27~pixels$  away along the north-south  direction, we avoided
the central  crowded portion of  the cluster. However, the  fields are
chosen in such a manner to include maximum number of member stars.  We
also  had a  large number  of sources  which are  not members  but are
present in the fields observed.  All the sources were manually checked
and  rejected  in case  of  an  overlapping.   But background  at  any
location of  the observed field  gets doubled due to  this overlapping
and may have significant effect if the background has variation within
$27~pixels$ (e.g., presence of reflection nebulosity). One star, BD+61
315, is  found to  be associated with  a faint nebulosity.  We suspect
that,  due to  the lack  of grid  in our  polarimeter,  the reflection
nebulosity may have contributed polarized light over the aperture used
for the photometry. We discuss this issue in section \ref{R_D}.

\section{Results \& Discussion}\label{R_D}

\begin{figure*}
\resizebox{12cm}{12cm}{\includegraphics{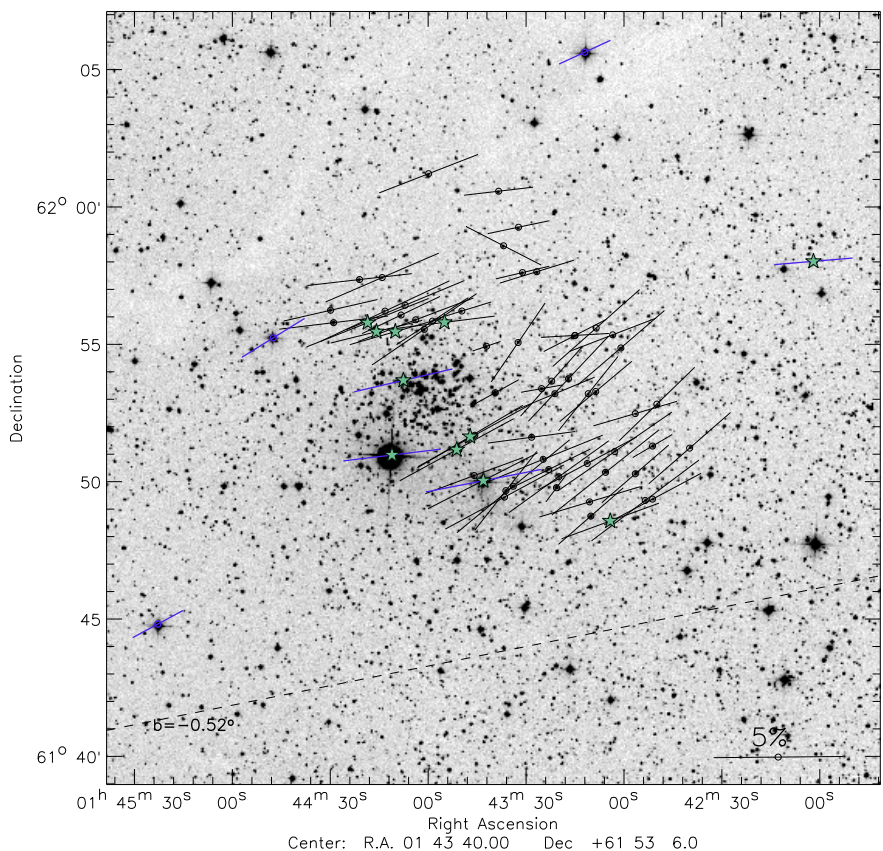}}
\caption{The  $28^{\prime}\times28^{\prime}$ R-band  DSS image  of the
field containing  NGC 654, reproduced from Digitized  Sky Survey.  The
position  angles, in  the equatorial  coordinate system,  are measured
from  the north,  increasing  eastward. The  polarization vectors  are
drawn with the  star as the center. Length  of the polarization vector
is proportional  to the percentage  of polarization $P_{V}$ and  it is
oriented  parallel  to the  direction  corresponding  to the  observed
polarization position angle $\theta_{V}$. A vector with a $P$ of $5\%$
is  shown  for reference.  The  dashed  line  represents the  Galactic
parallel   at  $b=-0.52^{\circ}$.   Stars  with   $M_{P}\geq0.70$  are
identified  with closed  star  symbols in  green colour.  Polarization
vectors of seven  stars observed by Pandey et al.  (2005) are shown in
blue colour.}
\label{pol_PA_DSSv.ps}
\end{figure*}

\begin{figure*}
\resizebox{12cm}{12cm}{\includegraphics{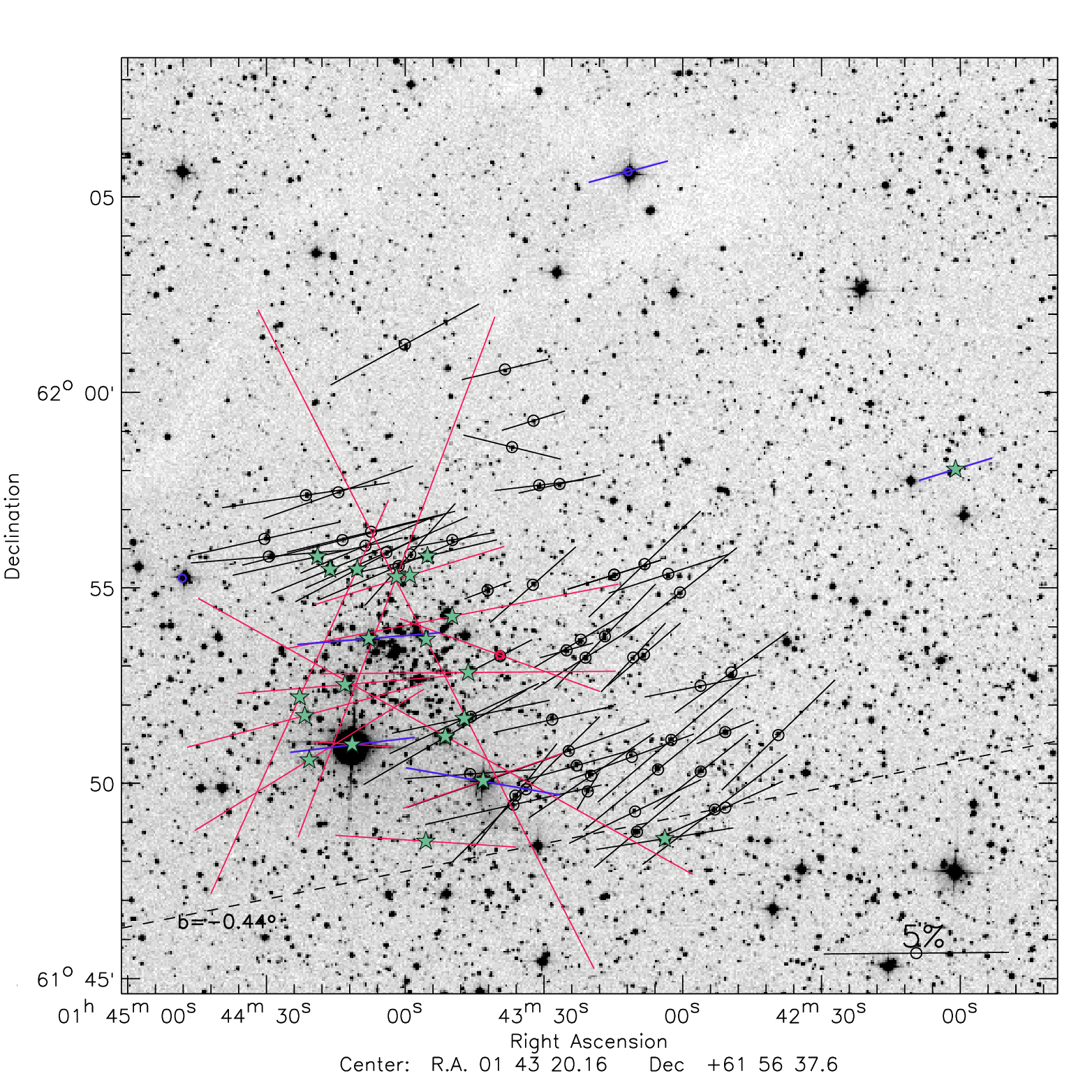}}
\caption{Same as in  Figure  \ref{pol_PA_DSSv.ps} but  for $P_{B}$ and
$\theta_{B}$.  Results for 14  stars observed  by Samson  (1976) using
IIa0  emulsion plates (which  has a  red cutoff  similar to  Johnson B
band) are  shown using vectors drawn  in red. Three out  of five stars
for which Pandey et al. (2005)  had B band measurements are also shown
using vectors in blue.}
\label{pol_PA_DSSb.ps}
\end{figure*}

Our polarimetric  results are  presented in the  Table \ref{result_1}.
Column 2  gives the  star identification as  given by Phelps  \& Janes
(1994).  Instrumental  magnitudes obtained in $V$ filter  are given in
column  3. The measured  values of  polarization $P$  (in \%)  and the
corresponding error $\epsilon$ (in \%) in  $B, V $ and $R$ filters are
given in columns 4, 6,  \& 8, respectively.  The polarization position
angle  (of the  ${\bf E}$  vector)  $\theta$ (in  $^{\circ}$) and  the
corresponding  error $\epsilon$  (in  $^{\circ}$) in  $B,  V$ and  $R$
filters are  given in  columns 5, 7  \& 9, respectively.  The position
angles in the equatorial coordinate system are measured from the north
increasing  eastward. Columns 10  \& 11  represent $E(B-V)$  (Joshi \&
Sagar 1983)  and the  membership probabilities ($M_{P}$)  (Stone 1977;
Joshi  \&   Sagar  1983;   Huestamendia,  Rio  \&   Mermilliod  1993),
respectively. Stars  with $M_{P}\geq  0.70$ are considered  as cluster
members in this study.

In Table \ref{lit_res},  we present previous polarization measurements
of stars in the direction of  NGC 645 carried out by Samson (1976) and
Pandey et al. (2005). Samson (1976) observed 14 stars in the direction
of NGC  654 using plates with  IIa0 emulsion. The IIa0  emulsion has a
red  cutoff similar  to Johnson  B  band. The  limiting magnitude  was
$B=15$ mag. In column 1,  we give the identification numbers which are
adopted from Samson (1976) and Pandey et al. (2005).  Columns 2, 3, 4,
5, 6,  7, 8,  \& 9 give  $P\%$ \& $\theta$  in B,  V, R \&  I filters,
respectively.   The errors  of measurements  are given  in parenthesis
wherever  available. Samson  (1976) gives  polarization  in magnitudes
($p$).   We converted  $p$ to  $P\%$ using  the  relation $P\%=46.05p$
(Whittet 1992).  The position angles  given by Samson (1976) are w.r.t
the east, as deduced from the  star \#54. We transformed the values to
w.r.t the north.  In column 10 we give the membership probabilities of
stars obtained from Stone (1977).

The sky projection of the V-band polarization vectors for the 61 stars
observed by us in NGC  654 are shown in Figure \ref{pol_PA_DSSv.ps} (R
band image is reproduced  from Digitized Sky Survey). The polarization
vectors are drawn  with the observed stars at  the center.  The length
of  the  polarization vector  is  proportional  to  the percentage  of
polarization in  V band ($P_{V}$) and  it is oriented  parallel to the
direction of  corresponding observed polarization position  angle in V
band ($\theta_{V}$). The dashed  line represents the Galactic parallel
at $b=-0.52^{\circ}$  inclined at $\sim100 ^{\circ}$  w.r.t the north.
The  stars with  $M_{P}\geq  0.70$ are  identified  using closed  star
symbols in green. Polarization vectors  for 7 stars observed by Pandey
et al.  (2005) are  also shown in  blue colour.  Clearly,  there exist
stars with  vectors (a) distributed  about the Galactic plane  and (b)
slightly greater than the  Galactic plane (especially those located to
the  west and  to the  south-western  regions of  the cluster).   This
indicates that  the dust  grains along the  line of sights  are mostly
aligned by a magnetic field  which is nearly parallel to the direction
of the Galactic  Disk. But a second component  of magnetic field which
is slightly inclined to the  Galactic Disk could also be present along
the line of sights of stars showing steeper angles.

\begin{figure}
\resizebox{8cm}{8cm}{\includegraphics{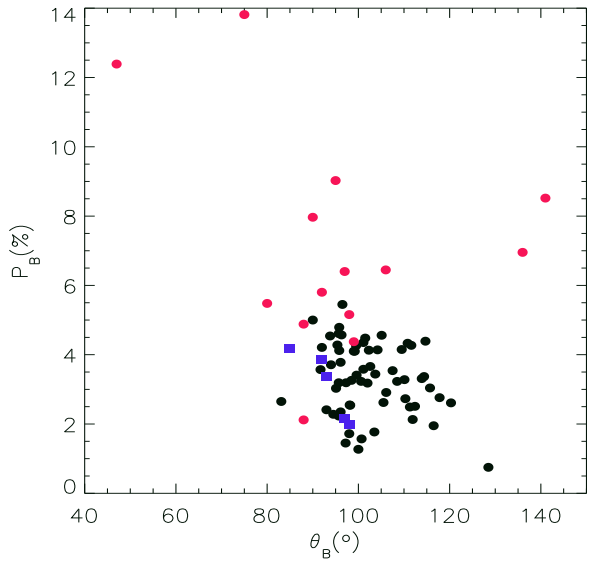}}
\caption{The $P_{B}$ vs. $\theta_{B}$ plot for 61 stars observed by us
in  the direction of  NGC 654  are shown  using black  filled circles.
Results of  stars observed by Samson  (1976) and Pandey  et al. (2005)
are shown using filled red circles and blue squares.}
\label{pol_PAb}
\end{figure}
\begin{figure}
\resizebox{8cm}{16cm}{\includegraphics{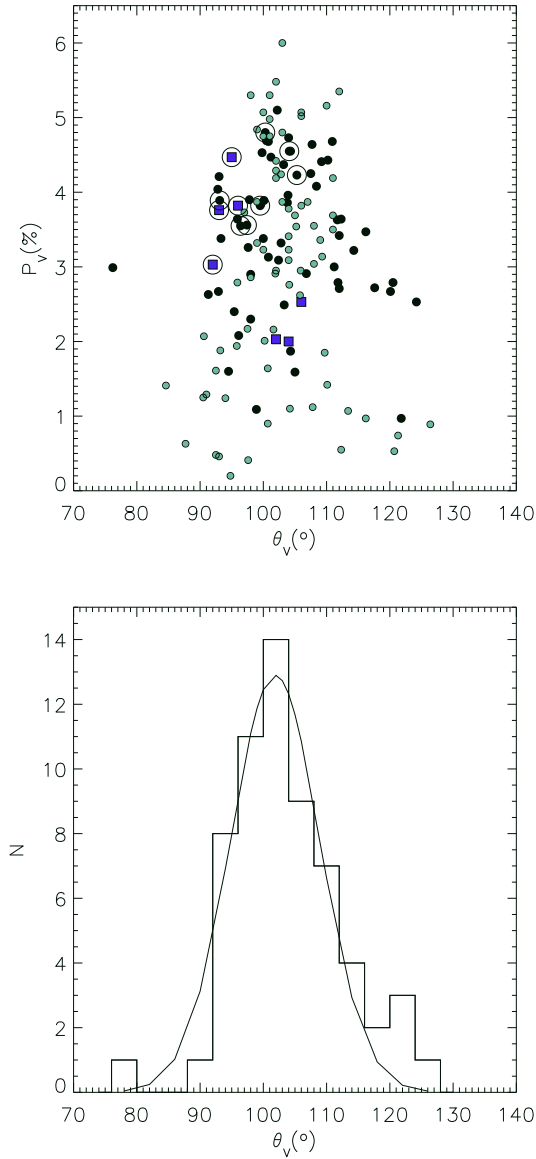}}
\caption{{\bf Upper  panel:} The $P_{V}$ vs. $\theta_{V}$  plot for 61
stars observed by us in the direction of NGC 654 are shown using black
filled circles.  Results of stars observed by Heiles (2000) and Pandey
et  al.  (2005)  are  shown   using  filled  green  circles  and  blue
squares.  The stars  with $M_p  \geq 0.70$  are identified  using open
circles.  {\bf Lower panel:}  Distribution of $V$ band position angles
for the stars from our observations}
\label{pol_PAv}
\end{figure}

In Figure  \ref{pol_PA_DSSb.ps} we present  the sky projection  of the
B-band polarization vectors for the 61 stars observed by us in NGC 654
(vectors in  black colour) along  with the results from  Samson (1976)
(in red) and Pandey et al. (2005)  (in blue). There are a few stars in
common among  the three observations.  The star  \#35, identified with
BD+61 315, from our list is the same as the star \#64 of Samson (1976)
and  the  star  \# 68  of  Pandey  et  al.   (2005). While  the  $P\%$
measurements  are  found  to  be  in agreement  among  all  the  three
observations  within  the uncertainty,  the  $\theta_B$  of Pandey  et
al. (2005) seems to be  $85^{\circ}$, $14^{\circ}$ less than the value
($99^{\circ}$)  obtained by  both in  our measurements  and  by Samson
(1976).  A faint  reflection nebulosity  (Kutner \etal  1980; Magakian
2003) is found to be associated  with BD+61 315.  We suspect that, due
to the lack of grid  in our polarimeter, the reflection nebulosity may
have  contributed  polarized light  over  the  aperture  used for  the
photometry. But,  since the nebulosity  is faint and no  stars located
along  the   north-south  except  one  close  to   BD+61  315  (within
$1^{\prime}$), we confidently believe that none of the other stars got
affected by this nebulosity. In addition to this, two other stars from
our  list,  stars \#34  \&  \#39,  are same  as  stars  \#56 \&  \#66,
respectively  in  Samson  (1976).  Both  of  them  showed  significant
differences  in their $P\%$  and $\theta_B$  when compared.   The star
\#66 from  Samson (1976) showed  the highest polarization  detected in
this direction  ($\sim14\pm0.5\%$). Inspection of another  star \#1 in
Samson (1976) which is same as the star \#111 in Pandey et al.  (2005)
showed a difference of  $\sim1\%$ polarization though $\theta_B$ seems
to be consistent within uncertainty.

In Figure \ref{pol_PAb}, we compare our results with those obtained by
Samson (1976) and Pandey et  al. (2005) using $P_{B}$ vs. $\theta_{B}$
plot. Our results are represented by filled black circles and those of
Samson  (1976) and Pandey  et al.  (2005) are  shown using  filled red
circles and blue squares, respectively. The results for stars observed
by us  are, in  general, consistent with  those observed by  Pandey et
al.  (2005). But  the $P_{B}$  obtained by  Samson (1976)  are showing
systematically  higher  values  (as  high as  $\sim14\%$)  than  those
obtained in both our observations and by Pandey et al. (2005).

We used  the Heiles (2000) catalogue  which has a  compilation of over
9000  polarization  measurements  to  determine whether  the  observed
polarization of  stars towards the NGC  654 by us  are consistent with
those available in the catalogue towards the same direction. For this,
we  searched the  catalogue for  stars with  polarization measurements
within  a circular  region of  radius $2^{\circ}$  around NGC  654. We
found 75  stars with polarization  measurements in $V$ band.  We could
not show  them in Figure  \ref{pol_PA_DSSv.ps} because the  nearest of
them      was      at     $\sim17^{\prime}$      ($01^{h}46^{m}03^{s},
+62^{\circ}01^{\prime}03^{\prime\prime}$)    away    from   NGC    654
($01^{h}44^{m}00^{s},  +61^{\circ}53^{\prime}06^{\prime\prime}$).   In
Figure  \ref{pol_PAv} upper  panel, we  show $P_{V}$  vs. $\theta_{V}$
plot. Our results are represented by filled black circles. The results
from Pandey et al.  (2005) and Heiles (2000) are represented by filled
green  circles and blue  squares. The  stars with  $M_{P}\geq0.70$ are
identified using open circles.

The $P_{V}$ of the  stars observed by us are found to  be in the range
of $\sim1$  to $\sim5\%$. The  mean value of $P_{V}$  and $\theta_{V}$
are found to be $3.5\pm1\%$ and $\sim104 \pm 9^{\circ}$, respectively.
The stars  from Heiles (2000)  show degree of polarization  ($P_H$) in
the range  from $\sim0.2$ to $\sim6$  \%. The mean value  of $P_H$ and
position  angle  are $\sim3\pm1.7$  \%  and  $\sim102^{\circ} \pm  9$,
respectively.  The range in degree of polarization and position angles
obtained       by       us       for      a       smaller       region
($\sim25^{\prime}\times25^{\prime}$) are similar to those for a larger
region from Heiles  (2000) implying that the cause  of polarization is
possibly due to  dust grains distributed in an  extended structure and
therefore likely to be located closer  to the Sun than the cluster.  A
number  of stars  selected from  Heiles (2000)  show higher  degree of
polarization  ($\gtrsim5$\%)  and  among  them, 5  stars  are  located
towards the direction of NGC 663 which is found to be at a distance of
$\sim2$ kpc (Kharchenko et al. 2005) similar to NGC 654. Thus stars in
this direction  can show degree  of polarization as high  as $\sim6\%$
but  certainly not  $\sim14\%$ as  obtained  by Samson  (1976). It  is
possible that  stars which showed  high polarization in  Samson (1976)
could  be  unique  and  definitely  require  further  multi-wavelength
investigations. Eight  out of 11 stars with  $M_{P}\geq0.70$ are found
to be clustering towards position angles which are lower than the mean
value of $104^{\circ}$.

In Figure  \ref{pol_PAv} lower  panel, we show  the distribution  of V
band position angles for stars from our observations. The distribution
is strongly peaked but shows a significant high position angle 'tail'.
A Gaussian fit to the  distribution is shown with peak value occurring
at 102$^{\circ}$  with a dispersion in position  angle of 7$^{\circ}$.
The dispersion in position angle is  found to be lower in NGC 654 when
compared  to  those for  NGC  6204  and  NGC 6193  (15.1$^{\circ}$  \&
16$^{\circ}$,  respectively,  Waldhausen \etal  1999)  and similar  to
those  for NGC  6611  (9.3$^{\circ}$, Bastien  \etal  2004), NGC  6167
(9.9$^{\circ}$,  Waldhausen  \etal  1999),  Stock  16  (6.7$^{\circ}$,
Feinstein \etal  2003) and IC 1805 (6.5$^{\circ}$,  Medhi \etal 2007).
Presence  of a  tail in  the distribution  of position  angles towards
other clusters (e.g., NGC 6167, Waldhausen \etal 1999) were attributed
to the  dust layers  located foreground to  those clusters.  In Figure
\ref{pol_PA_DSSv.ps}  \& \ref{pol_PA_DSSb.ps}  we noticed  presence of
opaque patches  to the north  of the cluster.  Therefore  to interpret
the  polarimetric   results,  it   is  important  to   understand  the
distribution of interstellar dust towards the direction of NGC 654.

\subsection{Distribution of interstellar matter in the region of NGC 654}

Samson (1975) noted that $E(B-V)$ for the stars in the central area of
NGC  654 were  lower than  the  stars located  near the  edges of  the
cluster. The radial increase in $E(B-V)$ was also noticed by Pandey et
al. (2005).  Samson (1975)  proposed the following three possibilities
to  explain  the  decrease  of  $E(B-V)$;  (i)  the  inner  stars  are
predominantly  cluster members  and lie  closer to  us than  the outer
ones, (ii) a  ring of obscuration lies between us  and the cluster and
(iii) the cluster  is embedded in a shell of  dust whose inner surface
has  a $3.6^{\prime}$  diameter.  He discarded  the first  possibility
because  of the  knowledge  of  distance modulus  to  inner and  outer
stars. The second one was discarded on the basis of statistical ground
arguing that  its unlikely that such  a ring of  obscuration would lie
exactly between us and NGC 654, thus favoring the third possibility of
a shell around NGC 645.

\begin{figure}
\resizebox{10cm}{9cm}{\includegraphics{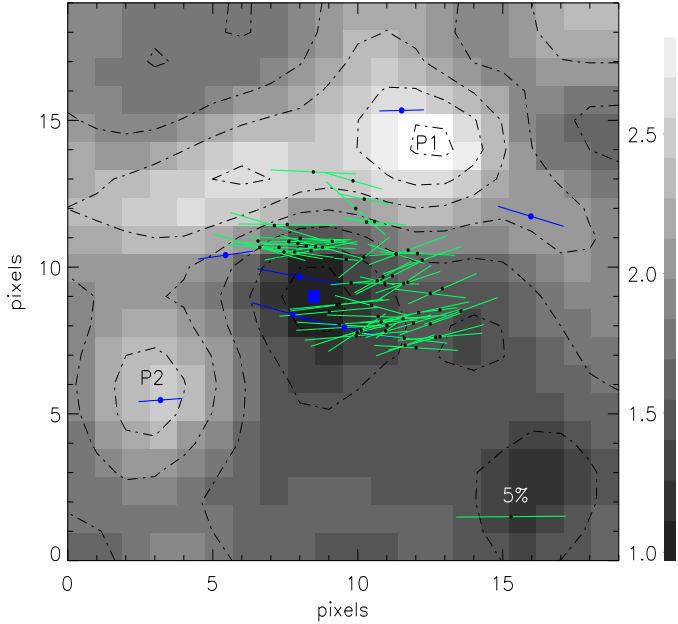}}
\caption{The   high   resolution   extinction   map  of   the   region
($30^{\prime}\times30^{\prime}$)  produced by  Dobashi  et al.  (2005)
using  the optical  database  `Digitized Sky  Survey  I' and  applying
traditional star count  technique. We overlay V band  results from our
observations  and from  Pandey et  al. (2005)  using vectors  drawn in
green  and  blue  colours,   respectively.  Colour  schemes  are  made
different  from previous  figures  for more  clarity.  The two  clumps
identified by Dobashi et al.  (2005) in this region are identified and
labeled as  P1 \& P2.  The center of the  cluster ($l=129.08^{\circ}$,
$b=-0.36^{\circ}$) is identified using blue square.}
\label{dobashi_v}
\end{figure}

\begin{figure}
\resizebox{10cm}{9cm}{\includegraphics{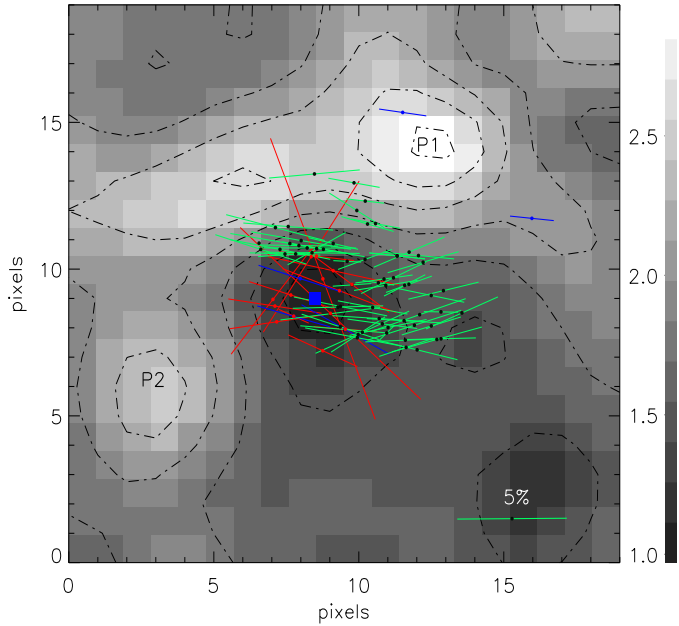}}
\caption{Same as the  Figure \ref{dobashi_v} but we overlay  with B band
results from  this work, Pandey et  al. (2005) and  Samson (1976). The
vectors shown in red colour are the results from Samson (1976).}
\label{dobashi_b}
\end{figure}

Dobashi et al. (2005) recently  produced extinction maps of the entire
region   of    the   Galaxy    in   the   galactic    latitude   range
$|b|\lesssim40^{\circ}$  using  the  optical database  `Digitized  Sky
Survey I'  and applying traditional star count  technique. We obtained
the fits images of the extinction  map of the field containing NGC 654
from their on-line website
\footnote{http://darkclouds.u-gakugei.ac.jp/astronomer/astronomer.html}.
In Figure  \ref{dobashi_v} we  present the high  resolution extinction
map overlaid with V band results from our observations and from Pandey
et  al.  (2005)  using  vectors  drawn  in  green  and  blue  colours,
respectively.    The  Figure   \ref{dobashi_b}  is   same   as  Figure
\ref{dobashi_v} but  is overlaid with  B band results from  this work,
Pandey  et al.  (2005) and  Samson (1976).   Because the  $A_{V}$ maps
shown are in galactic  coordinates, we transformed all position angles
measured relative to the equatorial  north to the Galactic north using
the relation given by Corradi et al. (1998). The square in blue colour
identifies   the   center    of   the   cluster   ($l=129.08^{\circ}$,
$b=-0.36^{\circ}$).  The colour-bar on  the right  shows the  range of
$A_{V}$ values in figures. The  contours are plotted at $A_{V}$=0.5 to
2.9 with an interval of 0.3 magnitude.

The extinction towards the location of the cluster (blue square) shows
relatively  low  ($A_{V}<1$) values.  But  the  outer  regions of  the
cluster  especially towards  the north  and the  east,  the extinction
increases up to $\sim3$ magnitude. Two clumps identified by Dobashi et
al.  (2005) in  these  regions are  labeled  as P1  \&  P2 in  Figures
\ref{dobashi_v} \&  \ref{dobashi_b}. Three dark clouds,  LDN 1332, LDN
1334  \& LDN  1337 identified  towards the  direction of  NGC  654 are
located close to P1 \& P2. While  LDN 1332 \& LDN 1334 are found to be
radially  $4^{\prime}$ \&  $7^{\prime}$, respectively,  away  from the
clump P1, LDN 1337 is found to be radially $11^{\prime}$ away from the
clump P2.   Note that the  $A_{V}$ values estimated in  the extinction
map  towards  the center  of  the cluster  are  less  than the  values
estimated by  Samson (1976) \&  Pandey et al.   (2005) by a  factor of
$\sim3$.  This  could be  because  of  the  coarse resolution  of  the
extinction map  by Dobashi et  al. (2005) and  the values could  be an
averaged one.

It  is interesting to  note that  the overall  morphology of  the dust
distribution in this region resembles  a ring structure with a central
hole  coinciding with the  position of  the cluster.  Structurally the
distribution  is highly  inhomogeneous. This  ring morphology  of dust
distribution explains the increase  in $E(B-V)$ values found by Samson
(1975) and Pandey et  al.  (2005) towards the edges  or outer parts of
the  cluster.  Therefore the  second  possibility  proposed by  Samson
(1976) which he discarded on a  statistical ground, arguing that it is
unlikely that a ring of  obscuration would lies exactly between us and
NGC 654,  seems to  be the  most plausible scenario.  We rule  out the
third possibility  proposed by Samson (1976), i.e.,  the cluster being
embedded in a shell of dust, because using star count method, it would
be difficult to detect dust  obscurations located at $\sim2$ kpc since
the  cloud would  become  inconspicuous  due to  the  large number  of
foreground stars.

\begin{figure}
\resizebox{9cm}{6cm}{\includegraphics{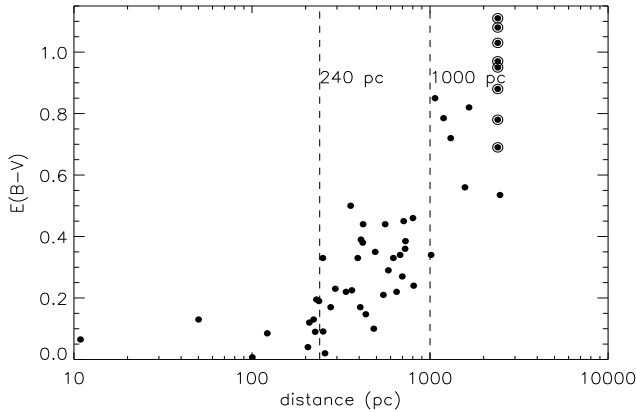}}
\caption{The distances of stars  are plotted against their E(B-V) from
the data obtained from Guarinos  (1992). The stars with $M_P\geq0.70$
are identified using open circles. These stars are assumed to be at
a distance of 2.4 kpc. The dotted lines are drawn at
240 pc and 1000  pc to show a sharp increase in  E(B-V) value at these
distances.}
\label{Guarinos_EBV}
\end{figure}

In order to determine distances to the foreground dust concentrations,
we  obtained distances  and $E(B-V)$  of main  sequence  stars located
within  a  circle  of  radius  $1^{\circ}$ around  NGC  654  from  the
catalogue \textit{Interstellar  matter in the  Galactic Disk} produced
by  Guarinos  (1992).   In  Figure  \ref{Guarinos_EBV},  we  show  the
$E(B-V)$ vs.   distance (pc) plot. The plot  shows $E(B-V)$ increasing
sharply at two distances, $\sim240$ pc and $\sim1$ kpc indicating that
their could  be at least  two layers of  dust one at $\sim240$  pc and
another at $\sim1$ kpc in the direction of NGC 654.  Using star counts
and counts  of extragalactic nebulae  Heeschen (1951) also  showed the
presence of  obscuring material at  $\sim200-300$ pc and  at $\sim800$
pc.   The obscuring material  located at  $\sim240$ pc  contributes an
$E(B-V)\sim0.5$ mag while net color  excess produced at 1 kpc is found
to  be $\sim0.85$  mag. The  stars with  $M_P\geq0.70$  are identified
using open circles. These stars are assumed to be at a distance of 2.4
kpc. From the figure it is evident that the extinction suffered by the
cluster members  of $\geq0.69$ mag are  mainly due to  the dust layers
present within 1 kpc from us.

In their photometric study of 23 open clusters, Phelps \& Janes (1994)
pointed out that the color-magnitude diagram of NGC 654 showed lack of
the characteristic wedge-shaped distribution  of the field stars.  The
cluster being located in front of a cloud is suggested to be the cause
(Phelps  \&  Janes 1994).  The  presence  of  a reflection  nebulosity
(Kutner \etal 1980; Magakian 2003)  associated with BD+61 315 (star \#
555, $M_P=0.84$)  shows that the  cluster is indeed associated  with a
cloud.  But, the relatively low $E(B-V)=0.88$ of this star, similar to
the $E(B-V)$  of 0.85  mag produced by  obscuring material  located at
$\sim200$ to 1 kpc (Figure \ref{Guarinos_EBV}), suggests that the star
(also  the  cluster) is  located  just  foreground  to the  associated
cloud.  This  is  further  supported  by  the  results  of  Pandey  et
al. (2005)  that at  the center of  the cluster $E(B-V)<1$  mag.  This
further  strengthen  the  argument  that the  extinction  towards  the
cluster is mainly contributed by the material distributed within 1 kpc
from  us.  Because  there is  a cloud  behind the  cluster,  the stars
including those with higher membership probability observed by us, may
not be  background to  the cloud behind  the cluster.   Henceforth, we
believe that the  observed polarization towards NGC 654  is due to the
aligned dust  grains associated with  the clouds located  at $\sim200$
and $\sim800-1000$ pc.

\begin{figure}
\resizebox{8cm}{14cm}{\includegraphics{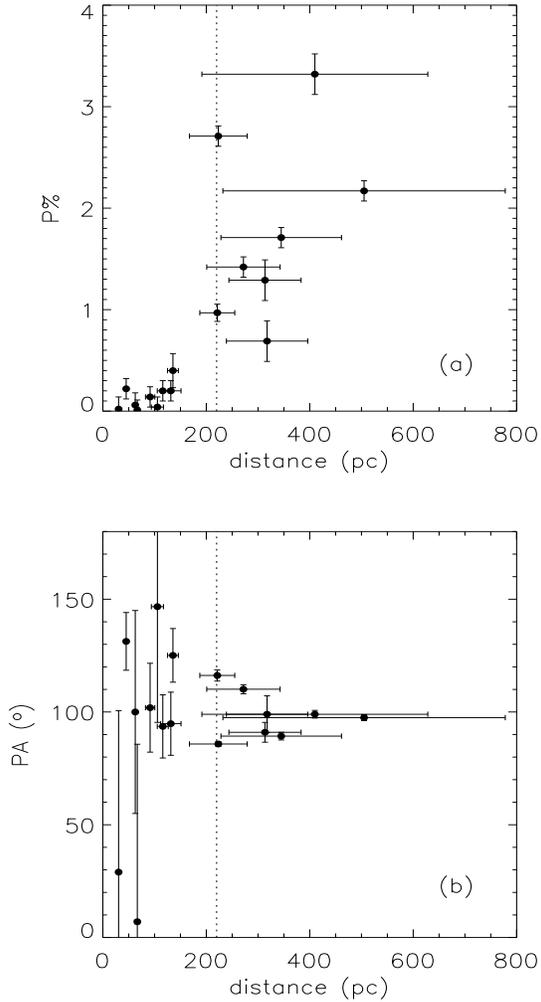}}
\caption{{\bf  (a)} The  polarization  as a  function  of distance  is
plotted.   The   distances  are  estimated   from  Hipparcos  parallax
measurements and  $P\%$ values are from Heiles  (2000) catalogue. {\bf
(b)} Position angles  are plotted for same set of  stars as a function
of their distances.}
\label{pol_dist}
\end{figure}

In  order to investigate the  polarization properties  of  dust grains
located at different distances from us, we selected all the stars from
within  a  circle of  radius  $3^{\circ}$  around  NGC 654  for  which
polarization and parallax measurements  are available in Heiles (2000)
and the  Hipparcos, respectively. The  Hipparcos parallax measurements
are  obtained  from  the   \textit  {Hipparcos  and  Tycho  Catalogs,}
(Perryman \etal 1997 and H$\phi$g \etal 1997).  We did not include the
stars observed by  us because of the lack  of distance information for
them except for the cluster member stars.  Stars which showed peculiar
features  and emissions  in their  spectrum, as  given by  SIMBAD, are
rejected.

In  Figures  \ref{pol_dist} (a)  and  (b)  we  present the  degree  of
polarization  ($P_{hip}$) and  position angle  ($\theta_{hip}$) versus
distance plots, respectively.  As expected, stars located closer to us
show  low $P_{hip}$ ($\lesssim0.5\%$)  with a  large scatter  in their
$\theta_{hip}$.  Stars located beyond  $\sim220$ pc  showed relatively
high values  of $P_{hip}$ in  the range from  $1\%$ to $3.3\%$  with a
sharp jump  in polarization ($\geq1\%$) occuring at  $\sim220$ pc. But
there are no  stars between $\sim150$ to $\sim200$ pc,  so we can only
infer  a maximum  distance  to  the dust  grains  responsible for  the
observed  sharp jump  from Figure  \ref{pol_dist}. However,  since the
abrupt  increase  in  both  $E(B-V)$ (Figure  \ref{Guarinos_EBV})  and
$P_{hip}$ (Figure \ref{pol_dist}) are  found to occur at $220-240$ pc,
we assign a maximum distance of $220$ pc to the first layer.

Among the two stars which showed  a sharp jump in $P_{hip}$ at $\sim220$
pc, the  one with higher  $P_{hip}$ (2.7\%) shows a  $\theta_{hip}$ of
$86\pm1^{\circ}$  and the  other with  lower $P_{hip}$  (1\%)  shows a
$\theta_{hip}$  of   $116\pm3^{\circ}$.   Beyond  $\sim220$   pc,  the
$P_{hip}$ increases with the  distance of the stars and $\theta_{hip}$
shows less scatter as it tend  to become more parallel to the Galactic
plane. This  implies that  though majority of  the dust grains  in the
direction of NGC 654 are aligned parallel to the Galactic plane, there
are  two additional  population of  dust grains  located in  the first
layer  at   $\sim  220$  pc   which  are  responsible   for  producing
polarization with position angles both  less than and greater than the
Galactic parallel.

If the polarization of the starlight is caused due to the alignment of
dust  grains with  the local  magnetic field,  then this  implies that
there exist two components of magnetic field which are responsible for
the  observed  polarization   properties.  While  the  magnetic  field
component aligned  more parallel with  the Galactic plane  is dominant
towards high  extinction regions,  the magnetic field  component which
are  found to  be steeper than  the  Galactic parallel  seems to  be
dominant towards the west and the south-west regions of NGC 654.

\subsection{Serkowski Law}
The maximum wavelength  ($\lambda_{max}$) and the maximum polarization
($P_{max}$)  both   are  functions  of  the   optical  properties  and
characteristic  of  particle size  distribution  of  the aligned  dust
grains (McMillan  1978; Wilking  et al. 1980).   Moreover, it  is also
related to  the interstellar  extinction law (Serkowski,  Mathewson \&
Ford 1975; Whittet \& van Breda 1978; Coyne \& Magalhaes 1979; Clayton
\&  Cardelli  1988).   The  $\lambda_{max}$ and  $P_{max}$  have  been
calculated  by fitting the  observed polarization  in the  B, V  and R
band-passes to the standard Serkowski's polarization law;

\begin{equation}
P_{\lambda}/    P_{max}   =\    exp    \left[-\   k    \   ln^{2}    \
(\lambda_{max}/\lambda) \right]
\end{equation}
and adopting the  parameter $k = 1.15 $ (Serkowski  1973). In the fits
the degree of freedom is adopted  as one.  Though there are only three
data points, the wavelength covered ranges  from 0.44 to 0.66 $\mu m $
and all the $\lambda_{max}$ found to fall within this range. Since, we
have  enough  wavelength coverage,  the  fit  is  reasonably fine  but
sometimes it may causes to over estimate the value of $\sigma_1$.  For
each star we  computed $\sigma_1$ parameter (the unit  weight error of
the fit).  If the polarization  is well represented by the Serkowski's
interstellar polarization  law, $\sigma_1$  should not be  higher than
1.6 due to  the weighting scheme.  A higher  value could be indicative
of the  presence of  intrinsic polarization.  The  $\lambda_{max}$ can
also give  us the  clue about the  origin of polarization.   The stars
which  have  $\lambda_{max}$  lower  than  the average  value  of  the
interstellar medium ($0.55 \pm 0.04  \ \ \mu m$, Serkowski \etal 1975)
are  the  probable  candidates  to  have  an  intrinsic  component  of
polarization (Orsatti, Vega and  Marraco 1998).  The dispersion of the
position angle ($\overline{\epsilon}$) for each star normalized by the
mean value of the position angle  errors is the another tool to detect
the  intrinsic  polarization.   The  values  obtained  for  $P_{max}$,
$\sigma_1$,  $\lambda_{max}$, and $\overline{\epsilon}$  together with
R.A.(2000J) and  Dec(2000J) for all  the 61 observed stars  with their
respective errors are given in Table 4.

Out of 61  observed stars three non-member stars  (namely, \#16, \#24,
and \#50) and one member star (namely, \#39) have the $\sigma_1$ value
above   the    limit   of   1.6.   Dispersion    in   position   angle
$\overline{\epsilon}$ is higher for the member star \#12 and nonmember
stars \#22, \#24, \#30, and  \#50.  Out of these above mentioned seven
probable candidates  detected by  using two criteria  $\sigma_1$ and
$\overline{\epsilon}$,  the value of  $\lambda_{max}$ is  smaller than
the  normal size  of the  grains  only for  the member  star \#12  and
nonmember star  \#16.  Although  the nonmember star  \#16 is  found to
have higher  value of $\sigma_1$  and lower value  of $\lambda_{max}$,
$\overline{\epsilon}$ is found to be  very small.  In the case of star
\#12, the  value of $\overline{\epsilon}$\ is high,  which indicates a
rotation  in  position angle  and  implies  that  the polarization  is
produced by a combination of different dust populations. Nevertheless,
in majority  of the  observed stars, the  polarization is found  to be
caused due to the foreground dust grains.

The  weighted mean  of $\lambda_{max}$  for the  member  and nonmember
stars are obtained as $ 0.52\pm 0.01 \ \mu m$ and $ 0.54\pm 0.01 \ \mu
m$ respectively.  There  are not much difference of  the weighted mean
of $\lambda_{max}$ between  the member and the nonmember  stars of NGC
654, which  imply that the lights  from the both  member and nonmember
stars   encountering   the   same   population  of   foreground   dust
grains. Moreover, these values are very close to the mean interstellar
value of  $\lambda_{max}$ 0.55 $\pm$  0.04 $\mu$m.  Therefore,  we can
also  conclude  that the  characteristic  grain  size distribution  as
indicated by the polarization study of stars in NGC 654 is nearly same
as that for the general interstellar medium.  The weighted mean of the
maximum polarization  for the  member stars and  non member  stars are
obtained as $ 3.74\pm 0.01 \ \%$ and $ 2.87\pm 0.01\ \%$ respectively.

\begin{figure*}
\resizebox{10cm}{10cm}{\includegraphics{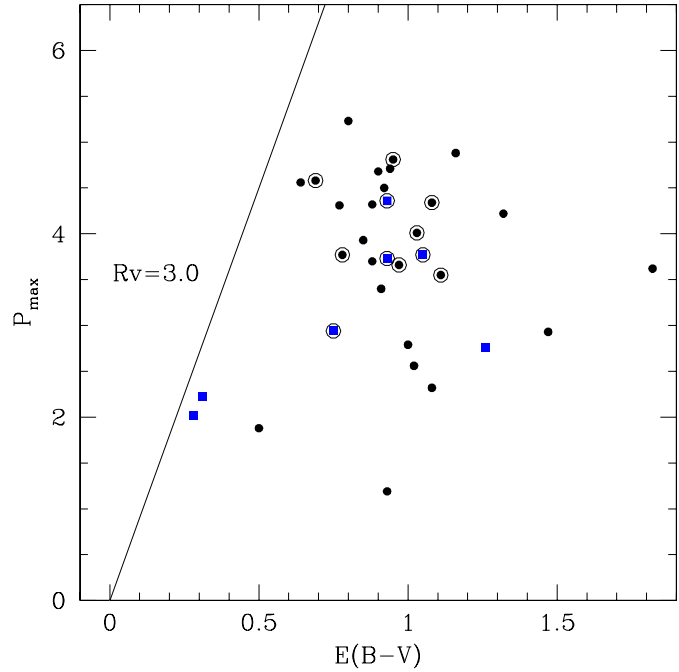}}
\caption{Polarization  efficiency diagram. Using  Rv=3.0, the  line of
maximum efficiency drawn.   The stars observed by us  in the direction
of NGC  654 are  shown using black  filled circles.  Results  of stars
observed  by  Pandey  et  al.  (2005)  are  shown  using  blue
squares.  The stars  with $M_p  \geq 0.70$  are identified  using open
circles.}
\end{figure*}

\subsection{Polarization efficiency}

For  interstellar dust  particles in  diffuse interstellar  medium the
ratio between the maximum  amount of polarization to visual extinction
(polarization  efficiency) can  not exceed  the empirical  upper limit
(Hiltner 1956),
\begin{equation}
\hspace*{16mm}{P_{max} < 3A_{V} \simeq 3R_{V} \times E(B-V)}
\end{equation}
The ratio $P_{max}/E(B-V)$ mainly depends on the alignment efficiency,
magnetic strength  and the amount  of depolarization due  to radiation
traversing more than one cloud in different direction.

Figure  9\  shows the  relation  between  colour  excess $E(B-V)$  and
maximum  polarization $P_{max}$ for  the stars  observed by  us (black
filled circle) and Pandey \etal (2005)(blue filled square) towards NGC
654  produced by  the  dust grains  along  the line  of  sight to  the
cluster. Colour excess $E(B-V)$ taken  from the work of Joshi \& Sagar
(1983); Huestamendia \etal  (1983). Among the 61 stars  observed by us
$E(B-V)$ is  available only for 26  stars (8 member  and 18 non-member).
In the efficiency  diagram none of the stars are lying  to the left of
the  interstellar maximum line.   In the  previous section,  though we
suspect  two member  stars \#12,  \#39  as a  candidate for  intrinsic
polarization by using different criterion,  they also fall to the left
in polarization efficiency  diagram.  From the polarization efficiency
diagram  it appears  that apparently  the  stars are  not affected  by
intrinsic polarization and the  dominant mechanism of polarization for
the observed  member of  NGC 654  is supposed to  be the  alignment of
grains  by magnetic  field,  like general  interstellar medium.   This
diagram also  indicates that  while the colour  excess for  the member
stars  of NGC  654 varies  from 0.69  to 1.11  mag  approximately, the
variation in the polarization value is high $\sim 1.87 \%$.  This high
variation  of $P_{max}$  indicates  the different populations of  dust
grains present in  the line of sight towards  NGC 654, as inferred from 
the Section 3.1.

\begin{figure*}
\resizebox{10cm}{10cm}{\includegraphics{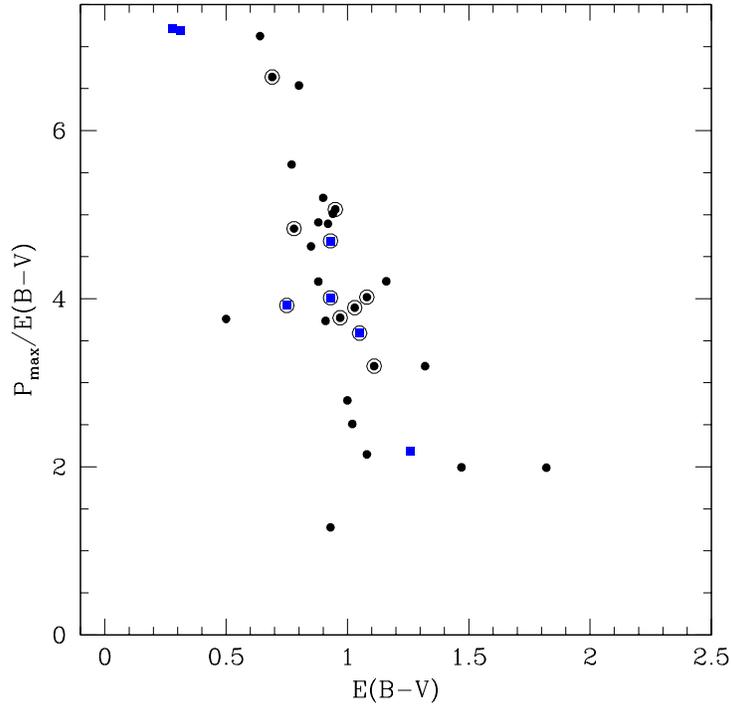}}
\caption{$P_{max}/E(B-V)$ plotted as a function of $E(B-V)$. The stars
observed  by us  in the  direction of  NGC 654  are shown  using black
filled circles.  Results of stars observed by Pandey et al. (2005) are
shown using  blue square. The  stars with $M_p  \geq 0.70$ are
identified using open circles. }
\end{figure*}

In Figure 10, we plot  the $P_{max}/E(B-V)$ vs. $E(B-V)$ for the stars
observed by  us and Pandey  \etal (2005) with available  colour excess
$E(B-V)$.  The  polarization efficiency  is  found  to  fall with  the
increase in $E(B-V)$.  There can  be two possibilities for this effect
(a)  the dust  grains  are located  in  local cloud  and  the drop  in
efficiency is due to increase in  the size of the grains; (b) the dust
grains  are  distributed along  the  line of  sight  and  the drop  in
efficiency  is due  to a  slight change  in the  polarization position
angle.

\section{Summary}

We have observed the linear polarization for 61 stars on the region of
the open  cluster NGC  654. The linear  polarization of the  stars are
caused due to the foreground  dust grains.  Combining our results with
those from the literature and various surveys, we present evidence for
the presence  of at least two dust  layers along the line  of sight to
the cluster.   Their distances are estimated  to be $\sim  200$ pc and
$\sim1$ kpc. Both dust layers  have their local magnetic field aligned
roughly parallel to the  Galactic plane. From the available extinction
maps, we show that the dust  towards NGC 654 are distributed in a ring
geometry  with  a  hole  which   coincides  with  the  center  of  the
cluster. This explains for an increase in the extinction towards outer
parts of the cluster inferred in previous studies of the region.

The  weighted mean of  maximum polarization  $P_{max}$ for  member and
nonmember stars are obtained as  $3.74\pm 0.01 \ \%$ and $2.87\pm 0.01
\  \%$,  respectively.   The   weighted  mean  of  maximum  wavelength
$\lambda_{max}$ for the cluster members  and the field stars are found
to  be  $0.52\pm   0.01  \  \mu  m$  and  $0.54\pm   0.01  \  \mu  m$,
respectively. These values of $\lambda_{max}$ of stars towards NGC 645
are thus  similar to those  of ISM ($0.55\pm  0.04 \ \mu  m$) implying
that  the polarization  is caused  mainly due  to the  foreground dust
grains as we have inferred for IC 1805 cluster (Medhi \etal 2007) also
in a previous study.

\section*{Acknowledgments}

We thank the referee for  his constructive remarks which lead to great
improvement in the  clarity of the paper.  This  research has made use
of the WEBDA database, operated  at the Institute for Astronomy of the
University  of  Vienna;  use   of  image  from  the  National  Science
Foundation and  Digital Sky  Survey (DSS), which  was produced  at the
Space Telescope  Science Institute under  the US Government  grant NAG
W-2166;  use of  NASA's  Astrophysics  Data System  and  use of  IRAF,
distributed  by  National Optical  Astronomy  Observatories, USA.  BJM
thanks Orchid for her support.

\section*{REFERENCES}
Bastien, P., Ménard, F., Corporon, P., Manset, N., Poidevin, F. \etal, 2004, Ap\&SS, 292, 427 \\	
Clayton, G. C., Cardelli, J. A., 1988, AJ, 96, 695 \\ 
Corradi, Romano L. M., Aznar, R., Mampaso, A., 1998, MNRAS, 297, 617 \\	
Coyne, G. V., Magalhaes, A. M., 1979, AJ, 84, 1200 \\
Davis, L. Jr., Greenstein, Jesse L., 1951, ApJ, 114, 206 \\
Dobashi, K., Uehara, H., Kandori, R., Sakurai, T., Kaiden, M., Umemoto, T., Sato, F., 2005, PASJ, 57, 1 \\
Feinstein, C., Baume, G., Vergne, M. M., Vazquez, R., 2003, A\&A, 409, 933 \\
Greenberg, J. M., 1968, Nebulae and Interstellar Matter, ed. B.M. Middlehurst \& L.H. Aller (Chicago Univ. Press), 221 \\
Guarinos J., 1992, Astronomy from Large Databases II", ESO Conference and Workshop Proceedings No 43, ISBN 3-923524-47-1, 301 \\
Heeschen, David S., 1951, ApJ, 114, 132 \\
Heiles, Carl., 2000, AJ, 119,923 \\
Hiltner, W. A., 1956, ApJS, 2, 389 \\
H$\phi$g, E., B\"{a}ssgen, G., Bastian, U., Egret, D., Fabricius, C., Gro$\beta$mann, V. \etal, 1997, A\&A, 323, 57 \\	
Huestamendia, G., del Rio, G., Mermilliod, J.C., 1993, A\&AS, 100, 25 \\
Joshi, U. C., Sagar, Ram, 1983, MNRAS, 202, 961 \\
Kharchenko, N. V., Piskunov, A. E., Röser, S., Schilbach, E., Scholz, R. D., 2005, A\&A, 440, 403 \\
Kutner, M. L., Machnik, D. E., Tucker, K. D., Dickman, R. L., 1980, ApJ, 237, 734 \\	
Lazarian, A., Goodman, Alyssa A., Myers, Philip C., 1997, ApJ, 490, 273 \\	
Lyng\.{a}\, G., 1984, Catalogue of open clusters Data,1/1S70401, Centre de Donne\.es Stellaires,Strasbourg. \\	
Magakian, T. Yu., 2003, A\&A, 399, 141 \\	
McMillan, R. S., 1978, APJ, 225, 880 \\
Medhi, Biman J., Maheswar, G., Brijesh, K., Pandey, J. C., Kumar, T. S., Sagar, R.,2007, MNRAS, 378, 881 \\	
Orsatti, A. M., Vega, E., Marraco, H. G., 1998, AJ, 116, 266 \\	
Pandey, A. K., Upadhyay, K., Ogura, K., Sagar, Ram, Mohan, V., Mito, H. \etal, 2005, MNRAS, 358, 1290 \\
Perryman, M. A. C., Lindegren, L., Kovalevsky, J., H$\phi$g, E., Bastian, U., Bernacca, P. L. \etal, 1997, A\&A, 323, 49 \\	
Phelps, Randy L., Janes, Kenneth A., 1994, ApJS, 90, 31 \\
Rautela, B. S., Joshi, G. C., Pandey, J. C., 2004, BASI, 32,159 \\
Sagar, Ram, Yu, Qian Zhong, 1989, MNRAS, 240, 551 \\
Samson, W. B., 1976, Ap\&SS, 44, 217 \\
Samson, W. B., 1975, Ap\&SS, 34, 363 \\	
Schmidt, G. D., Elston, R., Lupie, O. L., 1992, AJ, 104, 1563 \\
Serkowski, K., 1973, IAUS, 52, 145 \\
Serkowski, K., Mathewson, D. L., Ford, V. L., 1975, ApJ, 196, 261 \\
Stone, R. C.,1977, A\&A, 54, 803 \\ 
Waldhausen, Silvia, Martínez, Ruben E., Feinstein, Carlos, 1999, AJ, 117, 2882\\	
Whittet, D.C.B., van Breda, I.G., 1978, A\&A, 66, 57 \\
Whittet, D. C. B., 1992, Dust in the Galactic Environment (Bristol:IOP)\\
Wilking, B. A., Lebofsky, M. J., Kemp, J. C., Martin, P. G., Rieke, G. H., 1980, ApJ, 235, 905 \\

\begin{table*}
\centering
\begin{minipage}{650mm}
\caption{Observed $B, V$ and $R$ polarization values for different stars in NGC 654}\label{result_1}
\begin{tabular}{lllllllllll}
\hline
\hline
Sl.No.& Id(P) &V(mag) & $P_{B}$\pmi $\epsilon$ $(\%)$ & $\theta_{B}$ \pmi $\epsilon$ $(^{\circ})$ &  $P_{V}$ \pmi
$\epsilon$ $(\%)$ & $\theta_{V} $\pmi $\epsilon$ $(^{\circ})$& $P_{R}$\pmi  $\epsilon$ $(\%)$ & $\theta_{R} $
\pmi $\epsilon$ $(^{\circ})$& $E_{B-V}$ & $M_p$ \\
(1)&(2)&(3)&(4)&(5)&(6)&(7)&(8)&(9)&(10)&(11)\\
\hline
01 & \ ---     & 16.66   & 3.37 \pmi   0.56 &    114  \pmi    5  &  3.42  \pmi  0.25   &   112  \pmi   2  &   3.47  \pmi 0.27 &   108  \pmi    2 &\ --- &\ --- \\
02 & \ ---     & 16.16   & 3.23 \pmi   0.30 &    109  \pmi    3  &  3.22  \pmi  0.02   &   114  \pmi   2  &   3.20  \pmi 0.18 &   116  \pmi    2 &\ --- &\ --- \\
03 & \ ---     & 15.17   & 1.57 \pmi   0.08 &    101  \pmi    2  &  1.59  \pmi  0.05   &   105  \pmi   1  &   1.57  \pmi 0.07 &   101  \pmi    4 &\ --- &\ --- \\
04 & \ ---     & 17.05   & 3.18 \pmi   0.14 &    102  \pmi    1  &  3.86  \pmi  0.26   &   104  \pmi   2  &   3.35  \pmi 0.13 &   105  \pmi    2 &\ --- &\ --- \\
05 & \ ---     & 16.71   & 4.15 \pmi   0.65 &    110  \pmi    5  &  4.41  \pmi  0.13   &   109  \pmi   1  &   4.10  \pmi 0.53 &   112  \pmi    4 &\ --- &\ --- \\
06 & \ ---     & 14.96   & 2.73 \pmi   0.10 &    110  \pmi    1  &  2.79  \pmi  0.05   &   112  \pmi   1  &   2.49  \pmi 0.06 &   111  \pmi    1 &\ --- &\ --- \\
07 & \ ---     & 17.65   & 3.03 \pmi   0.37 &     95  \pmi    4  &  3.26  \pmi  0.40   &   98   \pmi   4  &   3.12  \pmi 0.07 &   101  \pmi    2 &\ --- &\ --- \\
08 & \ ---     & 15.99   & 3.28 \pmi   0.36 &    110  \pmi    3  &  3.63  \pmi  0.01   &   112  \pmi   1  &   3.49  \pmi 0.04 &   109  \pmi    1 &\ --- &\ --- \\
09 & \ ---     & 16.68   & 3.26 \pmi   0.49 &     99  \pmi    4  &  3.38  \pmi  0.28   &   100  \pmi   2  &   3.38  \pmi 0.28 &   100  \pmi    2 &\ --- &\ --- \\
10 & \ ---     & 17.35   & 3.31 \pmi   0.50 &    114  \pmi    4  &  3.64  \pmi  0.33   &   112  \pmi   3  &   3.79  \pmi 0.86 &   115  \pmi    7 &\ --- &\ --- \\
11 & \ ---     & 16.62   & 4.27 \pmi   0.13 &    112  \pmi    1  &  4.64  \pmi  0.27   &   108  \pmi   2  &   4.69  \pmi 0.12 &   112  \pmi    1 &\ --- &\ --- \\
12 &  703      & 13.11   & 3.71 \pmi   0.07 &     94  \pmi    1  &  3.82  \pmi  0.18   &   100   \pmi  2  &   3.41  \pmi 0.01 &   93   \pmi    1 & 0.78 & 0.70 \\
13 & \ ---     & 14.93   & 4.33 \pmi   0.04 &    111  \pmi    1  &  4.43  \pmi  0.05   &   110  \pmi   1  &   4.09  \pmi 0.17 &   108  \pmi    1 &\ --- &\ --- \\
14 & \ ---     & 14.74   & 2.49 \pmi   0.07 &    111  \pmi    1  &  2.71  \pmi  0.18   &   112  \pmi   2  &   2.39  \pmi 0.04 &   111  \pmi    1 &\ --- &\ --- \\
15 & \ ---     & 16.98   & 3.66 \pmi   0.37 &    103  \pmi    4  &  3.90  \pmi  0.64   &   98   \pmi   5  &   3.51  \pmi 0.29 &   104  \pmi    2 &\ --- &\ --- \\
16 & \ ---     & 16.96   & 3.41 \pmi   0.27 &     100  \pmi   4  &  3.96  \pmi  0.08   &   104  \pmi   1  &   3.52  \pmi 0.05 &   104  \pmi    1 &\ --- &\ --- \\
17 & \ ---     & 16.20   & 2.13 \pmi   0.36 &    112  \pmi    5  &  3.00  \pmi  0.45   &   111  \pmi   4  &   2.23  \pmi 0.11 &   109  \pmi    2 &\ --- &\ --- \\
18 & \ ---     & 17.28   & 1.95 \pmi   0.66 &    117  \pmi    10 &  2.67  \pmi  0.36   &   120  \pmi   2  &   2.25  \pmi 0.10 &   125  \pmi    1 &\ --- &\ --- \\
19 & \ ---     & 15.68   & 2.54 \pmi   0.18 &     98  \pmi    2  &  2.49  \pmi  0.21   &   103  \pmi   2  &   2.36  \pmi 0.11 &   97   \pmi    2 &\ --- &\ --- \\
20 & \ ---     & 16.08   & 2.55 \pmi   0.19 &     98  \pmi    2  &  2.63  \pmi  0.13   &   91   \pmi   1  &   2.47  \pmi 0.11 &   97   \pmi    2 &\ --- &\ --- \\
21 & \ ---     & 16.19   & 3.04 \pmi   0.09 &    116  \pmi    1  &  3.47  \pmi  0.25   &   116  \pmi   2  &   3.01  \pmi 0.19 &   109  \pmi    2 &\ --- &\ --- \\
22 & \ ---     & 15.71   & 2.91 \pmi   0.11 &    106  \pmi    1  &  3.13  \pmi  0.02   &   101  \pmi   1  &   2.98  \pmi 0.02 &   100  \pmi    1 &\ --- &\ --- \\
23 & \ ---     & 15.75   & 2.62 \pmi   0.33 &    106  \pmi    4  &  2.72  \pmi  0.20   &   118  \pmi   2  &   2.71  \pmi 0.24 &   109  \pmi    3 &\ --- &\ --- \\
24 & \ ---     & 15.86   & 1.45 \pmi   0.02 &     97  \pmi    1  &  1.60  \pmi  0.03   &   95   \pmi   1  &   1.33  \pmi 0.10 &   104  \pmi    2 &\ --- &\ --- \\
25 & \ ---     & 16.05   & 3.54 \pmi   0.38 &    108  \pmi    3  &  4.08  \pmi  0.34   &   108  \pmi   3  &   3.44  \pmi 0.10 &   107  \pmi    1 &\ --- &\ --- \\
26 & \ ---     & 15.54   & 0.75 \pmi   0.02 &    129  \pmi    1  &  0.97  \pmi  0.20   &   122  \pmi   6  &   0.79  \pmi 0.09 &   128  \pmi    4 &\ --- &\ --- \\
27 & \ ---     & 15.94   & 3.19 \pmi   0.10 &     97  \pmi    1  &  3.32  \pmi  0.30   &   103  \pmi   3  &   3.17  \pmi 0.03 &   102  \pmi    1 &\ --- &\ --- \\
28 & \ ---     & 17.33   & 4.26 \pmi   0.01 &     100  \pmi   1  &  4.69  \pmi  0.89   &   101  \pmi   5  &   3.90  \pmi 1.39 &   95   \pmi    10 &\ --- &\ --- \\
29 &  530      & 16.84   & 2.51 \pmi   0.39 &    112  \pmi    4  &  2.53  \pmi  0.03   &   124  \pmi   1  &   2.31  \pmi 0.48 &   121  \pmi    7 & 1.02 &\ --- \\
30 &  648      & 15.69   & 3.19 \pmi   0.06 &     96  \pmi    1  &  3.38  \pmi  0.04   &   93   \pmi   1  &   3.30  \pmi 0.03 &   91   \pmi    1 & 0.91 &\ --- \\
31 &  641      & 16.06   & 4.39 \pmi   0.03 &    115  \pmi    1  &  4.68  \pmi  0.26   &   111  \pmi   2  &   4.10  \pmi 0.17 &   116  \pmi    1 & 0.92 &\ --- \\
32 &  642      & 16.87   & 2.61 \pmi   0.44 &    120  \pmi    5  &  2.79  \pmi  0.07   &   121  \pmi   1  &   2.65  \pmi 0.20 &   121  \pmi    3 & 1.00 &\ --- \\
33 &  637      & 17.52   & 4.79 \pmi   0.41 &     96  \pmi    2  &  4.73  \pmi  0.22   &   104  \pmi   1  &   4.78  \pmi 0.12 &   100   \pmi   1 & 1.16 &\ --- \\
34 &  493      & 13.78   & 1.77 \pmi   0.01 &    104  \pmi    1  &  1.87  \pmi  0.01   &   104  \pmi   3  &   1.83  \pmi 0.01 &   100   \pmi   1 & 0.50 & 0.00 \\
35 &  555      & 10.74   & 4.10 \pmi   0.07 &     99  \pmi    1  &  4.37  \pmi  0.09   &   103  \pmi   1  &   4.12  \pmi 0.02 &   101  \pmi    1 & 0.88 & 0.84 \\
36 &  634      & 16.33   & 3.57 \pmi   0.12 &     92  \pmi    1  &  3.84  \pmi  0.26   &   100  \pmi   2  &   3.15  \pmi 0.13 &   98   \pmi    1 & 1.82 &\ --- \\
37 &  450      & 12.86   & 4.28 \pmi   0.31 &     95  \pmi    2  &  4.55  \pmi  0.03   &   104  \pmi   1  &   4.23  \pmi 0.22 &   102  \pmi    1 & 0.64 & 0.55 \\
38 &  439      & 14.53   & 4.13 \pmi   0.06 &    102  \pmi    5  &  4.23  \pmi  0.08   &   105  \pmi   1  &   4.17  \pmi 0.01 &   107  \pmi    2 & 1.08 & 0.89 \\
39 &  556      & 12.86   & 4.56 \pmi   0.11 &    105  \pmi    1  &  4.55  \pmi  0.03   &   104  \pmi   1  &   4.35  \pmi 0.01 &   105  \pmi    1 & 0.69 & 0.90 \\
40 & \ ---     & 15.67   & 2.23 \pmi   0.50 &     96  \pmi    6  &  2.90  \pmi  0.15   &   98   \pmi   2  &   2.48  \pmi 0.13 &   99    \pmi   2 &\ --- &\ --- \\
41 & \ ---     & 15.96   & 2.41 \pmi   0.29 &     93  \pmi    3  &  2.08  \pmi  0.36   &   96   \pmi   5  &   2.30  \pmi 0.08 &   99   \pmi    1 &\ --- &\ --- \\
42 & \ ---     & 16.53   & 2.65 \pmi   0.49 &     83  \pmi    5  &  2.99  \pmi  0.40   &   76   \pmi   3  &   2.43  \pmi 0.13 &   88  \pmi     2 &\ --- &\ --- \\
43 & \ ---     & 16.84   & 1.72 \pmi   0.35 &     98  \pmi    6  &  2.40  \pmi  0.56   &   95   \pmi   7  &   2.04  \pmi 0.11 &   98   \pmi    2 &\ --- &\ --- \\
44 & \ ---     & 16.38   & 2.35 \pmi   0.58 &     96  \pmi    7  &  2.67  \pmi  0.15   &   93   \pmi   3  &   2.54  \pmi 0.20 &   100  \pmi    2 &\ --- &\ --- \\
45 &  474      & 14.87   & 1.27 \pmi   0.20 &    100  \pmi    5  &  1.09  \pmi  0.11   &   99   \pmi   3  &   1.15  \pmi 0.04 &   94   \pmi    1 & 0.93 &\ --- \\
46 &  405      & 16.51   & 2.28 \pmi   0.10 &     95  \pmi    1  &  2.30  \pmi  0.09   &   98    \pmi  4  &   1.97  \pmi 0.25 &    103 \pmi    4 & 1.08 &\ --- \\
47 &  322      & 15.51   & 3.78 \pmi   0.17 &     96  \pmi    2  &  3.89  \pmi  0.26   &   93   \pmi   3  &   3.91  \pmi 0.16 &   95   \pmi    2 & 1.03 & 0.86 \\
48 &  267      & 16.72   & 2.76 \pmi   0.43 &    118  \pmi    5  &  2.91  \pmi  0.11   &   107  \pmi   1  &   2.62  \pmi 0.18 &   110  \pmi    2 & 1.47 &\ --- \\
49 & \ ---     & 16.19   & 4.14 \pmi   0.37 &    104  \pmi    3  &  3.89  \pmi  0.35   &   100  \pmi   1  &   3.83  \pmi 0.40 &   101  \pmi    3 &\ --- &\ --- \\
50 &  238      & 14.93   & 4.35 \pmi   0.20 &    101  \pmi    1  &  4.25  \pmi  0.04   &   108  \pmi   1  &   4.26  \pmi 0.01 &   101  \pmi    1 & 0.77 & 0.03 \\
51 &  187      & 16.20   & 4.48 \pmi   0.03 &    102  \pmi    1  &  4.53  \pmi  0.18   &   100   \pmi  1  &   4.44  \pmi 0.01 &   97   \pmi    2 & 0.90 &\ --- \\
52 &  150      & 17.26   & 4.57 \pmi   0.17 &     96  \pmi    1  &  4.68  \pmi  0.46   &   100  \pmi   3  &   4.40  \pmi 0.04 &   103  \pmi    1 & 0.94 & 0.11 \\
53 &  132      & 15.93   & 3.58 \pmi   0.12 &    101  \pmi    1  &  3.09  \pmi  0.48   &   102  \pmi   4  &   3.50  \pmi 0.09 &   97   \pmi    1 & 0.88 &\ --- \\
54 &  119      & 17.39   & 3.44 \pmi   0.61 &    104  \pmi    5  &  3.56  \pmi  0.62   &   97   \pmi   5  &   3.57  \pmi 0.15 &   94   \pmi    1 & 0.97 & 0.95 \\
55 &  080      & 17.18   & 5.45 \pmi   0.45 &     97  \pmi    11 &  5.10  \pmi  0.15   &   102  \pmi   2  &   4.80  \pmi 0.10 &   92   \pmi    2 & 0.80 & 0.11 \\
56 & \ ---     & 16.80   & 4.10 \pmi   0.69 &     99  \pmi    5  &  4.47  \pmi  1.08   &   101  \pmi   12 &   4.01  \pmi 0.31 &   102  \pmi    2 &\ --- &\ --- \\
57 &  061      & 16.63   & 3.23 \pmi   0.46 &    101  \pmi    15 &  3.55  \pmi  0.29   &   96   \pmi   1  &   3.46  \pmi 0.47 &   95   \pmi    4 & 1.11 & 0.83 \\
58 &  034      & 16.75   & 4.61 \pmi   0.61 &     96  \pmi    3  &  4.80  \pmi  0.17   &   100  \pmi   1  &   4.53  \pmi 0.10 &   99   \pmi    1 & 0.95 & 0.90 \\
59 & \ ---     & 15.74   & 4.54 \pmi   0.46 &     94  \pmi    3  &  4.04  \pmi  0.24   &   93   \pmi   2  &   4.12  \pmi 0.15 &   92   \pmi    1 &\ --- &\ --- \\
60 &  562      & 14.70   & 4.21 \pmi   0.20 &     92  \pmi    1  &  4.21  \pmi  0.03   &   93   \pmi   1  &   4.12  \pmi 0.15 &   92   \pmi    1 & 1.32 &\ --- \\
61 &  563      & 15.60   & 4.12 \pmi   0.55 &     96  \pmi    1  &  3.64  \pmi  0.34   &   96   \pmi   3  &   3.72  \pmi 0.07 &   93   \pmi    1 & 0.85 &\ --- \\
\hline
\end{tabular}
\end{minipage}
\end{table*}

\begin{table*}
\centering
\caption{Results of previous polarization measurements carried out 
by Samson (1976) and Pandey et al. (2005) in the direction of NGC 654. 
The errors in the measurements are given in parenthesis.}
\begin{minipage}{500mm}
\label{lit_res}
\begin{tabular}{llllllllll}
\hline
\hline
Id &$P_{B}$&$\theta_B$&$P_{V}$&$\theta_V$&$P_{R}$&$\theta_R$&$P_{I}$&$\theta_I$&$M_{P}$\\
   &(\%)&($^{\circ}$)&(\%)&($^{\circ}$)&(\%)&($^{\circ}$)&(\%)&($^{\circ}$)&(\%) \\
(1)&(2)&(3)&(4)&(5)&(6)&(7)&(8)&(9)&(10)\\
\hline
\multicolumn{10}{|c|}{Samson (1976)}\\
\hline
 1 &    2.12    (1.0)&  88& \ ---&\ --- &\ --- &\ --- &\ --- &\ --- &0.86\\
 5 &    6.45    (2.8)& 106& \ ---&\ --- &\ --- &\ --- &\ --- &\ --- &0.92\\
 8 &    6.95    (4.7)& 136& \ ---&\ --- &\ --- &\ --- &\ --- &\ --- &0.92\\
 9 &    6.40    (2.3)&  97& \ ---&\ --- &\ --- &\ --- &\ --- &\ --- &0.91\\
14 &    5.80    (2.9)&  92& \ ---&\ --- &\ --- &\ --- &\ --- &\ --- &0.82\\
36 &    5.16    (1.0)&  98& \ ---&\ --- &\ --- &\ --- &\ --- &\ --- &0.90\\
37 &    8.52    (0.4)& 141& \ ---&\ --- &\ --- &\ --- &\ --- &\ --- &0.92\\
40 &   12.39    (5.3)&  47& \ ---&\ --- &\ --- &\ --- &\ --- &\ --- &0.87\\
41 &    9.03    (1.2)&  95& \ ---&\ --- &\ --- &\ --- &\ --- &\ --- &0.82\\
54 &    7.97    (3.1)&  90& \ ---&\ --- &\ --- &\ --- &\ --- &\ --- &0.93\\
56 &    5.48    (1.7)&  80& \ ---&\ --- &\ --- &\ --- &\ --- &\ --- &0.00\\
64 &    4.37    (1.7)&  99& \ ---&\ --- &\ --- &\ --- &\ --- &\ --- &0.84\\
65 &    4.88    (2.0)&  88& \ ---&\ --- &\ --- &\ --- &\ --- &\ --- &0.89\\
66 &   13.82    (0.5)&  75& \ ---&\ --- &\ --- &\ --- &\ --- &\ --- &0.90\\
\hline
\multicolumn{10}{|c|}{Pandey et al. (2005)}\\
\hline
9  &2.00(0.47)&98(7)&3.03(0.16)&92(2) &2.68(0.11)&94(1) &2.31(0.17)&98(2) &0.92\\
57 &2.16(0.36)&97(5)&2.03(0.15)&102(2)&1.48(0.14)&98(3) &1.25(0.26)&97(5) &0.90\\
68 &4.18(0.70)&85(5)&4.47(0.19)&95(1) &3.78(0.11)&94(1) &3.42(0.18)&95(2) &0.84\\
109&3.86(0.22)&92(2)&3.82(0.06)&96(1) &3.23(0.05)&96(1) &3.00(0.07)&95(1) &0.90\\
111&3.37(0.26)&93(2)&3.76(0.09)&93(1) &3.32(0.06)&93(1) &2.78(0.08)&93(1) &0.86\\
137&\ ---   &\ ---  &2.53(0.32)&106(4)&2.03(0.18)&114(3)&1.51(0.29)&111(6)&0.00\\
161&\ ---   &\ ---  &2.00(0.21)&104(3)&1.94(0.17)&110(2)&1.81(0.29)&103(5)&0.48\\
\hline
\end{tabular}
\end{minipage}
\end{table*}
\begin{table*}
\centering
\begin{minipage}{110mm}
\caption{The $P_{max}$, $\lambda_{max}$, $\sigma_1$ \& $\overline{\epsilon}$ for the observed data in NGC 654.}
\begin{tabular}{llllllll}
\hline
\hline
Sl.No.& R.A(2000J) & DEC(2000J)&{$P_{max}$\pmi $\epsilon$} $(\%)$ & $\sigma_1$  & $\lambda_{max}$\pmi$\epsilon$ ($\mu m$)&$\overline{\epsilon}$ \\
\hline
01 & 01\ 42\ 39.54 & 61 \ 51\ 18.4  & 3.50 \pmi 0.08 & 0.43  &  0.58  \pmi0.05&0.81  \\
02 & 01\ 42\ 49.68 & 61 \ 52\ 50.9  & 3.22 \pmi 0.02 & 0.83  &  0.56  \pmi0.05&1.32  \\
03 & 01\ 42\ 51.02 & 61 \ 51\ 19.4  & 1.62 \pmi 0.01 & 0.74  &  0.54  \pmi0.03&0.92  \\
04 & 01\ 42\ 51.00 & 61 \ 49\ 23.4  & 3.49 \pmi 0.15 & 1.55  &  0.57  \pmi0.05&0.56  \\
05 & 01\ 42\ 53.27 & 61 \ 49\ 20.5  & 4.41 \pmi 0.13 & 0.20  &  0.53  \pmi0.02&0.44  \\
06 & 01\ 42\ 56.33 & 61 \ 50\ 18.7  & 2.81 \pmi 0.05 & 0.93  &  0.48  \pmi0.02&1.06  \\
07 & 01\ 42\ 56.36 & 61 \ 52\ 30.2  & 3.23 \pmi 0.01 & 0.08  &  0.55  \pmi0.01&0.66  \\
08 & 01\ 43\ 00.81 & 61 \ 55\ 00.8  & 3.63 \pmi 0.01 & 0.42  &  0.55  \pmi0.01&0.58  \\
09 & 01\ 43\ 03.34 & 61 \ 55\ 29.5  & 3.44 \pmi 0.05 & 0.26  &  0.57  \pmi0.03&0.22  \\
10 & 01\ 43\ 08.45 & 61 \ 55\ 44.4  & 3.71 \pmi 0.07 & 0.15  &  0.61  \pmi0.03&0.21  \\
11 & 01\ 43\ 02.69 & 61 \ 51\ 07.4  & 4.74 \pmi 0.12 & 0.25  &  0.60  \pmi0.01&2.21  \\
12 & 01\ 43\ 04.08 & 61 \ 48\ 35.7  & 3.77 \pmi 0.03 & 0.59  &  0.49  \pmi0.01&4.35  \\
13 & 01\ 43\ 05.56 & 61 \ 50\ 21.9  & 4.45 \pmi 0.11 & 0.29  &  0.51  \pmi0.01&1.78  \\
14 & 01\ 43\ 10.11 & 61 \ 48\ 46.3  & 2.57 \pmi 0.04 & 0.86  &  0.51  \pmi0.02&0.33  \\
15 & 01\ 43\ 10.51 & 61 \ 49\ 17.4  & 3.79 \pmi 0.05 & 0.22  &  0.51  \pmi0.02&0.64  \\
16 & 01\ 43\ 11.22 & 61 \ 50\ 41.3  & 3.93 \pmi 0.24 & 2.14  &  0.49  \pmi0.05&1.09  \\
17 & 01\ 43\ 08.59 & 61 \ 53\ 18.5  & 2.39 \pmi 0.02 & 1.46  &  0.52  \pmi0.13&0.28  \\
18 & 01\ 43\ 10.91 & 61 \ 53\ 13.8  & 2.42 \pmi 0.03 & 0.95  &  0.52  \pmi0.12&0.72  \\
19 & 01\ 43\ 15.02 & 61 \ 55\ 28.1  & 2.57 \pmi 0.03 & 0.28  &  0.50  \pmi0.01&1.20  \\
20 & 01\ 43\ 15.03 & 61 \ 55\ 28.2  & 2.64 \pmi 0.01 & 0.03  &  0.52  \pmi0.01&1.63  \\
21 & 01\ 43\ 17.03 & 61 \ 53\ 46.4  & 3.22 \pmi 0.11 & 1.09  &  0.55  \pmi0.04&2.01  \\
22 & 01\ 43\ 21.23 & 61 \ 53\ 13.6  & 3.13 \pmi 0.01 & 0.71  &  0.54  \pmi0.01&4.63  \\
23 & 01\ 43\ 22.23 & 61 \ 53\ 41.2  & 2.76 \pmi 0.04 & 0.27  &  0.56  \pmi0.02&1.68  \\
24 & 01\ 43\ 25.32 & 61 \ 53\ 24.8  & 1.57 \pmi 0.01 & 2.29  &  0.57  \pmi0.05&3.23  \\
25 & 01\ 43\ 20.00 & 61 \ 50\ 12.4  & 3.79 \pmi 0.27 & 1.11  &  0.50  \pmi0.06&0.26  \\
26 & 01\ 43\ 20.74 & 61 \ 49\ 48.3  & 0.82 \pmi 0.05 & 0.78  &  0.58  \pmi0.05&0.84  \\
27 & 01\ 43\ 23.13 & 61 \ 50\ 28.0  & 3.33 \pmi 0.01 & 0.04  &  0.54  \pmi0.01&1.77  \\
28 & 01\ 43\ 24.85 & 61 \ 50\ 50.4  & 4.45 \pmi 0.19 & 0.36  &  0.54  \pmi0.05&0.45  \\
29 & 01\ 43\ 32.36 & 61 \ 55\ 12.7  & 2.56 \pmi 0.01 & 0.05  &  0.50  \pmi0.01&1.15  \\
30 & 01\ 43\ 28.43 & 61 \ 51\ 38.5  & 3.40 \pmi 0.01 & 0.49  &  0.56  \pmi0.01&3.49 \\
31 & 01\ 43\ 34.08 & 61 \ 49\ 51.6  & 4.50 \pmi 0.16 & 0.89  &  0.51  \pmi0.02&2.00 \\
32 & 01\ 43\ 36.39 & 61 \ 49\ 41.9  & 2.79 \pmi 0.01 & 0.13  &  0.54  \pmi0.01&0.15\\
33 & 01\ 43\ 36.81 & 61 \ 49\ 27.9  & 4.88 \pmi 0.12 & 0.94  &  0.57  \pmi0.04&1.80 \\
34 & 01\ 43\ 39.68 & 61 \ 53\ 15.9  & 1.88 \pmi 0.01 & 1.53  &  0.56  \pmi0.01&1.56 \\
35 & 01\ 43\ 43.51 & 61 \ 50\ 03.5  & 4.32 \pmi 0.05 & 0.64  &  0.54  \pmi0.02&1.42 \\
36 & 01\ 43\ 46.18 & 61 \ 50\ 14.9  & 3.62 \pmi 0.12 & 1.26  &  0.48  \pmi0.03&2.17 \\
37 & 01\ 43\ 46.03 & 61 \ 51\ 42.9  & 4.56 \pmi 0.15 & 0.48  &  0.53  \pmi0.02&2.70 \\
38 & 01\ 43\ 47.45 & 61 \ 51\ 40.1  & 4.34 \pmi 0.04 & 1.41  &  0.55  \pmi0.08&0.66 \\
39 & 01\ 43\ 51.62 & 61 \ 51\ 11.3  & 4.58 \pmi 0.05 & 1.78  &  0.53  \pmi0.01&1.13 \\
40 & 01\ 43\ 26.83 & 61 \ 57\ 40.2  & 2.84 \pmi 0.13 & 1.46  &  0.48  \pmi0.10&0.36  \\
41 & 01\ 43\ 31.24 & 61 \ 57\ 38.3  & 2.41 \pmi 0.15 & 0.98  &  0.53  \pmi0.07&0.67  \\
42 & 01\ 43\ 37.10 & 61 \ 58\ 36.8  & 2.83 \pmi 0.28 & 0.74  &  0.47  \pmi0.06&1.46  \\
43 & 01\ 43\ 32.48 & 61 \ 59\ 17.4  & 2.06 \pmi 0.08 & 0.71  &  0.62  \pmi0.09&0.23  \\
44 & 01\ 43\ 38.64 & 62 \ 00\ 36.0  & 2.65 \pmi 0.04 & 0.30  &  0.55  \pmi0.03&0.65  \\
45 & 01\ 43\ 42.38 & 61 \ 54\ 57.3  & 1.19 \pmi 0.08 & 1.17  &  0.55  \pmi0.01&0.93\\
46 & 01\ 43\ 50.02 & 61 \ 56\ 14.3  & 2.32 \pmi 0.04 & 0.52  &  0.49  \pmi0.02&1.00 \\
47 & 01\ 43\ 55.40 & 61 \ 55\ 49.5  & 4.01 \pmi 0.05 & 0.51  &  0.56  \pmi0.07&0.57 \\
48 & 01\ 43\ 59.12 & 61 \ 55\ 52.2  & 2.93 \pmi 0.07 & 0.40  &  0.49  \pmi0.03&1.68 \\
49 & 01\ 44\ 00.30 & 62 \ 01\ 14.3  & 4.11 \pmi 0.15 & 0.58  &  0.49  \pmi0.04&0.84  \\
50 & 01\ 44\ 01.63 & 61 \ 55\ 34.3  & 4.31 \pmi 0.05 & 2.46  &  0.59  \pmi0.03&4.16 \\
51 & 01\ 44\ 04.29 & 61 \ 55\ 55.1  & 4.68 \pmi 0.11 & 0.78  &  0.53  \pmi0.01&1.66 \\
52 & 01\ 44\ 07.58 & 61 \ 56\ 26.9  & 4.71 \pmi 0.01 & 0.02  &  0.52  \pmi0.01&1.84 \\
53 & 01\ 44\ 08.81 & 61 \ 56\ 05.5  & 3.70 \pmi 0.10 & 1.27  &  0.53  \pmi0.03&1.32 \\
54 & 01\ 44\ 10.59 & 61 \ 55\ 29.3  & 3.66 \pmi 0.01 & 0.17  &  0.57  \pmi0.07&0.97 \\
55 & 01\ 44\ 13.77 & 61 \ 56\ 14.1  & 5.23 \pmi 0.17 & 0.88  &  0.50  \pmi0.03&0.71 \\
56 & 01\ 44\ 14.67 & 61 \ 57\ 27.4  & 4.28 \pmi 0.08 & 0.20  &  0.52  \pmi0.02&0.16  \\
57 & 01\ 44\ 16.44 & 61 \ 55\ 29.1  & 3.55 \pmi 0.02 & 0.08  &  0.58  \pmi0.08&0.35 \\
58 & 01\ 44\ 19.17 & 61 \ 55\ 48.8  & 4.81 \pmi 0.06 & 0.05  &  0.53  \pmi0.01&0.96 \\
59 & 01\ 44\ 21.67 & 61 \ 57\ 23.0  & 4.29 \pmi 0.27 & 1.42  &  0.53  \pmi0.08&0.33  \\
60 & 01\ 44\ 29.70 & 61 \ 55\ 48.9  & 4.22 \pmi 0.04 & 1.14  &  0.53  \pmi0.04&0.43 \\
61 & 01\ 44\ 30.64 & 61 \ 56\ 15.9  & 3.93 \pmi 0.26 & 1.04  &  0.53  \pmi0.07&0.87 \\
\hline
\end{tabular}
\end{minipage}
\end{table*}


\bsp
\label{lastpage}
\end{document}